# Plasmonic Nanobubbles – A Perspective

Seunghyun Moon[1], Qiushi Zhang[1], Zhihao Xu[1], Dezhao Huang[1], Seongmin Kim[1], Jarrod Schiffbauer[2,*], Eungkyu Lee[3,*], Tengfei Luo[1,4,*]

1. Department of Aerospace and Mechanical Engineering, University of Notre Dame, IN, USA
2. Department of Physical and Environmental Sciences, Colorado Mesa University, Co, USA
3. Department of Electronic Engineering, Kyung Hee University, Yongin-si, South Korea
4. Department of Chemical and Biomolecular Engineering, University of Notre Dame, IN, USA

* Corresponding authors: jschiffbauer@coloradomesa.edu; eleest@khu.ac.kr; tluo@nd.edu.

**Abstract**:

The field of plasmonic nanobubbles, referring to bubbles generated around nanoparticles due to plasmonic heating, is growing rapidly in recent years. Theoretical, simulation and experimental studies have been reported to reveal the fundamental physics related to this nanoscale multi-physics phenomenon. Using plasmonic nanobubbles for applications is in the early stage but progressing. In this article, we briefly review the current state of this research field and give our perspectives on the research needs in the theoretical, simulation and experimental fronts. We also give our perspectives on how the fundamental understanding can be applied to more practical applications.

## 1. Introduction:

Due to the surface plasmon resonance (SPR), plasmonic nanoparticles (NPs) can efficiently convert photon energy into heat when excited by light at the SPR frequency. Such intensely heated NPs can locally generate the so-called plasmonic nanobubbles. These nanobubbles are known for their unique photothermal and optical properties and have already led to biomedical applications in cell-level therapy and imaging, controlled drug release and delivery, microtissue surgery and biosensing, with some already entered clinic trials.[1-5] They are also studied for energy and fluidic applications like solar-vapor generation,[6,7] plasmon-assisted photocatalytic reactions,[8] optofluidics,[9] nano swimmers,[10] surface bubble manipulation[11] and materials assembly.[12] In case the plasmonic NPs are immobilized or fabricated on a substrate, they can form bubbles on the surface upon optical excitation, and we refer to them as plasmonic surface bubbles, but our focus in this article is on plasmonic nanobubbles, which are formed around NPs suspended in liquids.

Besides the promises in applications, the fundamentals of plasmonic nanobubbles are no less attractive. The physics involved in this molecular-to-nanometer scale phenomenon is complicated, and studies have tried to understand it from different angles. The respective roles of surface chemistry, curvature, viscosity, and surface tension in bubble formation and subsequent dynamics have been at the focus of theoretical and numerical investigations of plasmonic nanobubbles for over a decade,[13-17] but are not fully resolved. The inherently multi-scale thermodynamics of phase change at nanoscale-curved solid-fluid interfaces poses several challenges from a theoretical viewpoint. The $1/r$ (where $r$ is the radius of NP) interfacial contributions to the free energy, surface-solvent interactions, solvent properties, and competition between time-scales of NP heating and cooling all play important roles in the nucleation and subsequent dynamics of plasmonic nanobubble, leading to effects such as an experimentally observed minimum in the threshold fluence,[14,16] explosive bubble collapse and oscillations,[18,19] and reported NP surface temperatures above the melting point of gold.[20] Several theoretical approaches have been brought to bear on these problems,[19,21-24] elucidating the underlying complexities of the problem in doing so, especially for bubble nucleation and dynamics for NPs with radii of tens of nm and smaller. However, in general, these theoretical approaches have relied on simplified fluid models for a single species of the molecule, while the complexities of experimentally relevant multi-component fluids are often ignored.

In this perspective, we briefly review the current state of the field of plasmonic nanobubbles, but instead of a comprehensive review, we focus on what we believe to be fundamentally interesting and give our perspectives on the challenges and unresolved questions in the theoretical, simulation, and experimental fronts. We will emphasize the fundamentals of plasmonic nanobubbles but will also touch on their application aspects. Besides, we also briefly discuss plasmonic surface bubbles when they are related to plasmonic nanobubbles.

## 2. State of the art and Perspective:

Two of the main questions that have driven much of the fundamental research on plasmonic nanobubbles are: (1) how are nanobubble generated upon optical excitation? and (2) what are the bubble dynamics? Answering these questions is no easy task. On the one hand, the nanometer bubble size and generally sub-nano to picosecond dynamics makes the experimental study of nanobubbles challenging, while on the other hand, the complex interplay of fluid transport properties and surface forces in confined or highly curved geometries makes unraveling the mechanisms underlying these issues a difficult task from a theoretical perspective. In the following sections, we will discuss the state of addressing each of these questions.

**Plasmonic nanobubble formation:** Theoretically, much of the focus has been on the role of NP diameter, laser pulse duration ($\tau$), and Kapitza resistance (*i.e.*, interfacial thermal resistance between NP and solvent)[23] on the fluence threshold for plasmonic nanobubble formation. However, the dissipative properties of the fluid and thermodynamics of the NP are also found critical in determining the heat flux from the hot particle to surrounding fluid.[24, 25] Metwally et al.[23] demonstrated that, for pulsed heating of NPs in a simple, uniform solvent, the origins of the minimum in the fluence threshold for NPs with radius of $r \sim 20\text{-}30$ nm lie chiefly in a competition between the timescales of the electron-photon interaction and that of diffusive cooling. At NP radii below this, diffusive losses to the environment require increasingly higher fluence for the NP surface to reach the spinodal temperature of the solvent, with larger losses at longer pulse durations. With increasing radius much above 20-30 nm, damping of the absorption cross-section results in a deviation from its linear dependence on the NP volume, leading to an increase in the fluence threshold (**Fig. 1a**). For sufficiently large NPs, the diffusive cooling time will generally be larger than the thermal relaxation time associated with the interfacial thermal conductance, thus Metwally et al.[23] conclude that interfacial thermal conductance plays a negligible role in the determination of threshold fluence. However, because the variation of thermal properties and phase of the surrounding fluid are not accounted for, the applicability their conclusions regarding the role of interfacial thermal conductance may be restricted to nanostructures with sufficiently large radii of curvature, i.e., accordingly for $r \gg \kappa/G$, where $\kappa$ is the solvent thermal conductivity, and $G$ the interfacial thermal conductance. Thus, as a practical example, for gold-water interfaces, this implies that NPs with radii of curvature much smaller than ~4 nm require further consideration (shaded area in **Fig. 1a**), but such small NPs are not often seen in applications. However, as shown by Wang et al.,[26] accounting for the full thermodynamics of the NP, interfacial thermal conductance, pulse duration, and wavelength ($\lambda$) can become critical factors, shifting the overall threshold fluence curve upward around the 20-80 nm range (in water) and yielding criteria for optimally "biosafe" NP bubble nucleation. At radii much below or above, the competition between dissipation and absorption overshadows the role of structural/phase change in the NP.

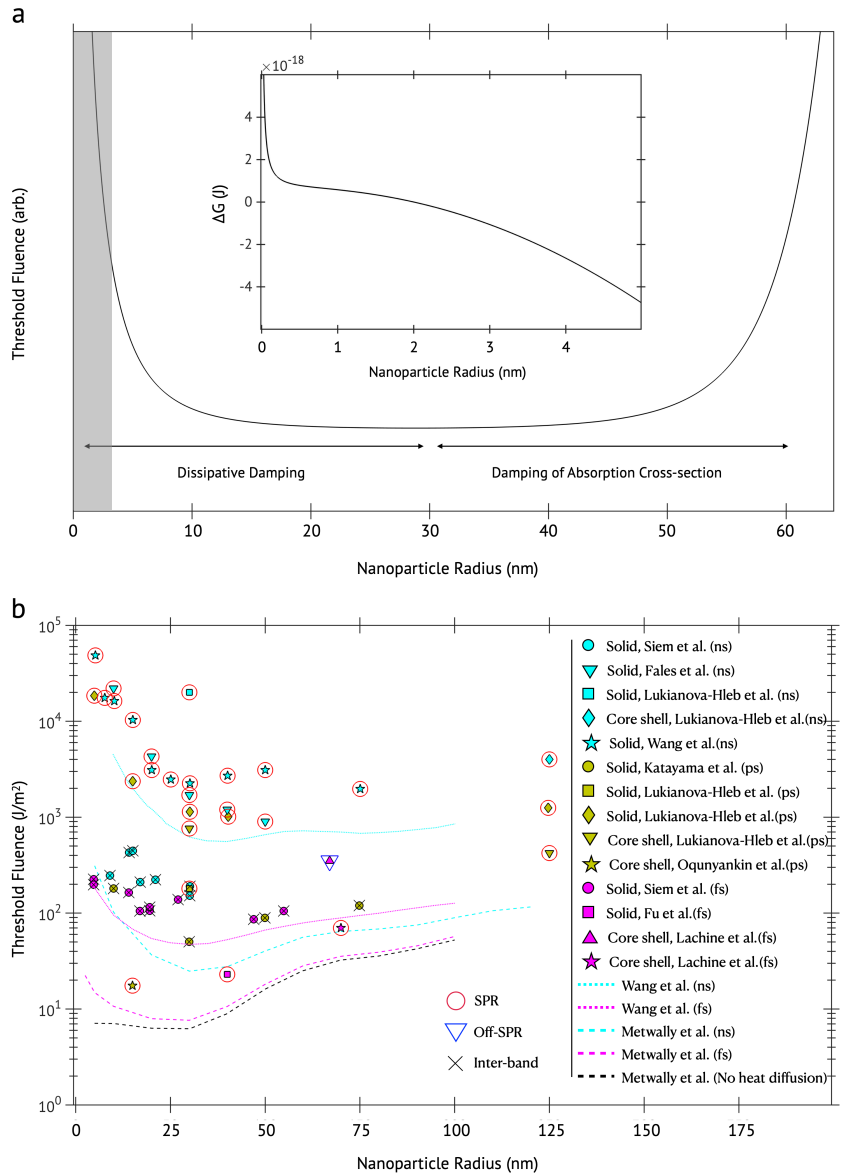

**Figure 1**. (a) Schematic of threshold fluence vs. NP radius for general pulsed heating cases showing the ubiquitous "bathtub" minimum determined by the competition between heating and cooling timescales. At sufficiently small radius (grey shaded region), the formation of vapor becomes interface-controlled and the detailed properties of the fluid and the wetting properties of the interface must be considered along with the thermodynamics of the NP itself. This is explained by the inset showing the Gibbs energy change for nucleation of a ~1 nm-thick vapor layer around a moderately hydrophilic NP from liquid Argon in thermal equilibrium at the liquid Argon spinodal. For NPs smaller than 2 nm in this fluid, no stable bubble can be formed, in agreement with the non-equilibrium MD results of Sasikumar et al.[22] (b) Nanobubble formation fluence threshold as a function of radius of Au NP for different laser pulse durations. Legends describe the type of NP, reference and laser pulse duration in parenthesis: Solid, Siem et al. (ns);[14] Solid, Fales et al. (ns);[27] Solid, Lukianova-Hleb et al. (ns);[16] Core-shell, Lukianova-Hleb et al. (ns);[13] Solid, Wang et al. (ns);[26] Solid, Katayama et al. (ps);[15] Solid, (square) Lukianova-Hleb et al. (ps);[16] Solid, (diamond) Lukianova-Hleb et al. (ps);[13] Core shell, Lukianova-Hleb et al. (ps);[13] Core shell, Oqunyankin et al. (ps);[28] Solid, Siem et al. (fs);[14] Solid, Fu et al. (fs);[29] Core-shell, Lachaine et al. (fs);[30] (Dotted Lines) Wang et al.,[26] (Dash lines) Metwally et al.[23]

Overall, the competition between heating and cooling timescales for NPs with different sizes would lead to a ubiquitous "bathtub" minimum in the fluence threshold for nanobubble formation (**Fig. 1a**). This trend can generally explain experimental observations from different studies. **Figure 1b** summarizes the fluence thresholds for plasmonic nanobubble formation around Au NPs with various sizes reported in the last decade. Most of the experiments have used solid spherical Au NPs immersed in water since they can be easily synthesized and tuned in size. In **Fig. 1b**, we can have the following key observations that shed light on different influential thermal and optical behaviors of the system of interest.

First, smaller Au NPs usually have higher fluence thresholds in each experimental setup. As the NP size gets smaller, it has a smaller thermal mass and larger surface-to-volume ratio, which renders more efficient heat dissipation to the solvent and faster cooling of the NP. Higher fluence is then required to balance the faster cooling to reach the bubble formation threshold. The predictions by the models from Metwally et al. (dash lines)[23] and Wang et al. (dotted lines)[26] clearly show such threshold increments for small NPs very well. Without considering the enhanced heat diffusion for such NPs, Metwally's model would show that the increased fluence threshold no longer exists (black dash line). On the other extreme, when the NP size gets large, its thermal mass increases, and the optical absorption per volume usually decreases. Thus, the fluence threshold will also increase, which can also be predicted correctly by the models. The crossover of these two opposite trends leads to the ubiquitous "bathtub" with a minimum in the fluence threshold as a function of NP size.

Second, the fluence threshold differ for different excitation laser pulse duration. The fluence influences the total energy that can be deposited in a pulse, and the pulse duration controls the photon-electron energy transfer rate. An interesting observation in **Fig. 1b** is that for the picosecond or nanosecond pulse durations, the optical excitation at the inter-band transition frequency (near 330 nm or 470 nm)[31] enables the bubble generation at a lower fluence than that at the SPR frequency. This is not intuitive because, at the SPR, the free electron gas in the NP strongly interacts with the incident photon, leading to stronger optical absorption and thus more intense heating than that at the inter-band transition frequency. As a result, the SPR excitation should have a lower fluence threshold than that of the inter-band excitation. This is the prediction by Metwally et al.[23] for SPR excitation (cyan dash line) which is one order of magnitude smaller than those of experimental sets with the inter-band excitation (cyan circle, $\tau = 10$ ns, $\lambda = 355$ nm: inter-band). However, the measured threshold fluences at the SPR excitation differ from the model prediction[23] by two orders of magnitude. One possible reason for the difference is that the SPR is sensitive to the change of refractive indices of the system. When the pulse duration is comparable to the electron-phonon relaxation time (1~2 ps),[23] the thermal energy can be released from the hot electrons during the laser absorption process, which increases the temperature of NP and medium (e.g., water) and thus changes their refractive indices. Indeed, Wang et al.[26] found that medium (*e.g.*, surrounding water) heating can suppress the SPR effect. This is also known as SPR bleaching, which limits NP's ability to fully leverage the incident photon in a pulse. When Wang et al.[26] consider the temperature-dependent optical absorption efficiency of NP in their model, the predicted threshold fluence (cyan dotted line, $\tau = 5$ ns, $\lambda = 530$ nm: SPR) becomes closer to the experimental result, but they still differ by one order of magnitude. The reason of this discrepancy remains an open question to be answered. On the other hand, the inter-band transition is not sensitive to temperature change, as it is related to the band-to-band transition of an electron by absorbing a photon. As a result, the optical absorption will not be significantly affected in the photo-thermal heating process, which may have led the inter-band excitation threshold fluence to be lower than the SPR cases. However, using femtosecond pulsed laser can avoid the SPR bleaching effect. The characteristic time of medium heating by the cooling of Au NP is ~100 ps,[32] which is longer than the electron-phonon relaxation time, thus femtosecond pulses can finish the interaction with electrons before the medium heating. Indeed, Fu et al.[29] used a 4-dimensional transmission electron microscope to visualize the nanobubble generation around solid Au NPs, and found that a femtosecond pulse at the SPR with a fluence of 23 Jm$^{-2}$ could form a nanobubble on the NP (pink square, $\tau$ = 350 fs, $\lambda$ = 520 nm: SPR). This reported value is close to the predicted threshold fluence by Metwally et al.[23] (pink dash line). Also, Fu's threshold fluence is much smaller than the measured fluence with inter-band transition excitation with a femtosecond pulse (pink circle and pink dotted line, $\tau$ = 100 fs, $\lambda$ = 400 nm: inter-band). Therefore, it may be preferential to use femtosecond pulsed laser to leverage the SPR effect in plasmonic nanobubble applications.

It is worth noting that Metwally et al.[23] emphasize the water spinodal temperature (~550 K) as the correct thermodynamic criterion for liquid-vapor transition, rather than referring to the critical temperature. But in the presence of highly curved surfaces, and accounting for the role of surface forces and wettability on the thermodynamics of phase change at the interface, as well as the interfacial thermal conductance,[33, 34] the liquid spinodal itself can shift.[35, 36] The equilibrium and non-equilibrium thermodynamics of phase change near heated NP interfaces has been studied using both molecular dynamics [21, 22] and continuum/phase-field methods.[24] Both liquid-vapor and solid-liquid interfacial energetic costs must be met to create a stable bubble. However, in general, the solid-vapor interfacial energy cost is higher. This is both owing to the larger surface energy per unit area required to create a dry surface, especially for a hydrophilic interface, as well as the fact that the inner NP radius is smaller than the radius at the liquid-vapor interface. Rather than resulting in a dry interface (**Fig. 2a**), for hydrophilic surfaces, this fact leads to the stabilization of a superheated liquid layer adjacent to the NP surface (**Fig. 2b**);[21, 22, 24] the energetic cost of forming an interface with such a layer will necessarily be lower than that of a dry interface. This phenomenon has been observed in MD simulations,[37] which show a layer of liquid molecules adhered to the NP surface despite that a vapor bubble is formed when the NP is sufficiently heated (**Fig. 2c and 2d**). However, the thermodynamic properties of this thin liquid layer, including its Gibbs energy at equilibrium, are presumably quite distinct from that of the liquid bulk or the usual liquid-vapor interface. While an equilibrium Gibbs energy criterion for a "dry" NP-bubble[24] (see **Fig. 1a** inset) can lead to estimates for equilibrium and critical NP radii for bubble nucleation in good quantitative agreement with simulations of hydrophilic NPs, as well as estimates for equilibrium bubble size for a given NP, it is not yet clear that the thermodynamic properties of this liquid layer can be included in such a formulation in a straightforward way. Non-equilibrium continuum thermodynamics can capture the stabilization of the superheated liquid layer.[24]

It is worth noting that while the photothermal phase change interpretation remains to be the main stream to understand the plasmonic nanobubble formation, Lachine et al.[30] proposed a very different mechanism for nanobubble formation. They have shown that at off-SPR wavelengths, the Au NP can still form a nanobubble, but concluded that it is due to the intense electric field near the NP, where the induced local electric field can overcome the optical breakdown threshold to gasify water.

When determining the fluence threshold, a prerequisite is the ability to detect nanobubble formation. Since the formation of nanobubbles can change the scattered light intensity, optical pump-probe transient scattering methods have been used to detect nanobubble formation. The pump beam is the excitation laser, while the probe beam, much weaker in energy than the pump beam, can be set at the wavelength where the variation of scattering efficiency is sensitive to nanobubble formation.[10] Since nanobubble formation is accompanied by an acoustic wave, the acoustic response has also been used for its detection.[26, 27] Direct observation of nanobubble formation has also become possible when a time-resolved transmission electron microscope is employed.[29]

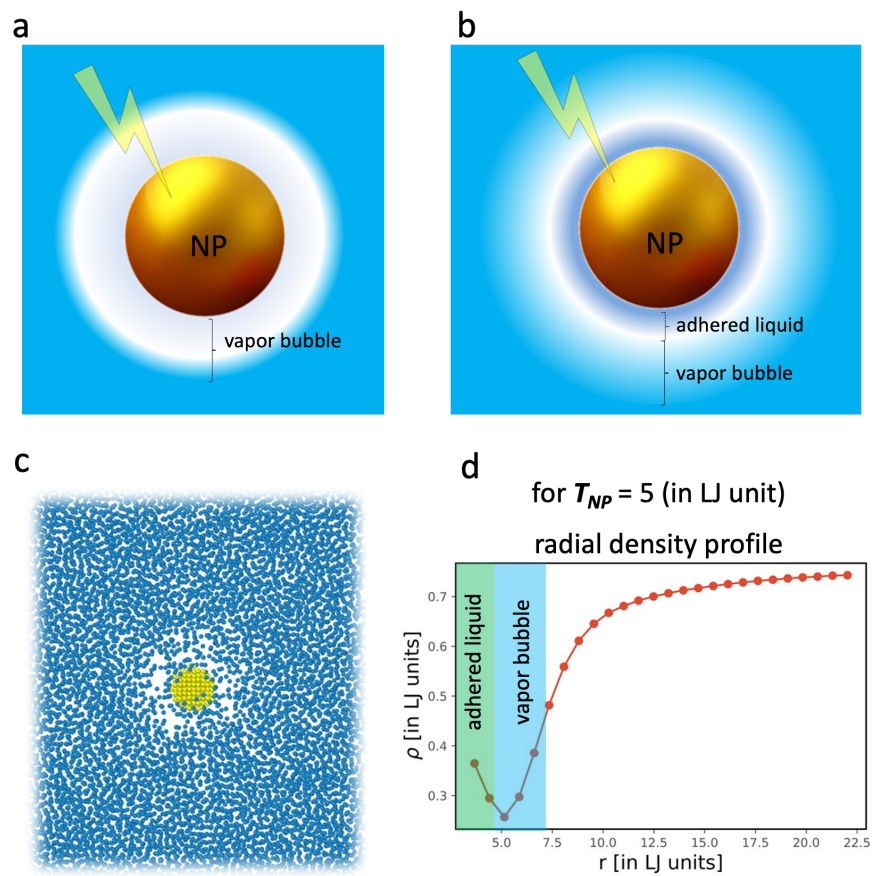

**Figure 2**. Schematics of (a) a "dry" bubble and (b) a bubble with an adhered liquid layer on the NP surface. MD simulation snapshot showing an adhered liquid layer (c), which is quantified by the radial density profile (d) for NP heated to $T_{NP}$ = 5 in LJ (Lennard-Jones) unit.

**Nanobubble dynamics:** The bubble dynamics, lifetime, and eventual collapse involve more detailed consideration of the properties of the fluid and fluid-vapor interface. Phase-field models and the Rayleigh-Plesset equation have been shown to yield comparable bubble dynamics including collapse and oscillations,[18, 38] where the dynamics depend on the heating condition (hence thermodynamics) of the NP.[19] Sasikumar and Keblinski[38] identified four distinct stages of nanobubble formation: (1) nucleation and adiabatic expansion of hot vapor, (2) isothermal expansion, (3) isothermal collapse, and (4) rapid heating. The range of times for bubble dynamics depends somewhat on the NP size, as well as the properties, and composition of the fluid. However, allowing for these differences, Sasikumar and Keblinski's MD simulations[38] agree well with the experimental observations and continuum simulations of Kotaidis and Plech,[18] with the initial adiabatic expansion occurring during the first ~50-100 ps, followed by expansion and collapse over the next 100 ps to 1 ns.

Recent experimental studies[39, 40] demonstrate that physicochemical factors such as dissolved gas or other solutes, especially as they modify the surface tension and viscosity, are important, particularly in the dynamics of bubble collapse. However, in these studies, it is not clear how the thermodynamic and physicochemical properties of the solvent itself, independent from the changes in effective thermophysical properties due to the NPs themselves, affect bubble dynamics. One often ignored consideration in modeling and theory is the role of dissolved gas at the liquid-vapor interface, which has been recently shown[40] to have a profound effect on bubble collapse dynamics. In addition to contributions to the bulk solvent properties, dissolved gasses can severely impact the structure and composition of liquid-vapor and fluid-solid interfaces, and thereby play an important role in both nanobubble nucleation and collapse dynamics. Overall, the issue of the role of dissolved gases on interfacial behavior in fluids is complex and not well understood. Through both ionic and non-ionic mechanisms, ranging from simple steric contributions to interfacial structure to possible induced dipole or even Casimir effects,[41] dissolved gasses can dramatically alter the balance and nature of interfacial forces governing capillary phenomena. Thus, an important direction for further work in plasmonic-generated bubbles can include detailed studies of the non-equilibrium thermodynamics of multi-component solutions and the role of high curvature/specific area.

**NP dynamics with nanobubbles:** Beyond bubble nucleation and subsequent dynamics, another issue of importance is the possible motion of an NP within a nanobubble. Here, it is worth noting that, while the problem can be treated in the framework of driven Brownian motion, the near-particle gradients in temperature, density, and viscosity can lead to complexities in the determination of the effective friction,[42] and such effects become more localized and dominate NP dynamics when a nanobubble is formed to encapsulate it (*i.e.*, supercavitating NP). Huang et al.[37] found that an intensely heated supercavitating NP would exhibit ballistic Brownian motion with the effective friction similar to that in a gas (**Fig. 3a**). The key is that the NP is kept hot in the nanobubble so that it can instantaneously evaporate water as it moves to keep itself in a gaseous environment – a nanoscale analogy of the Leidenfrost effect (**Fig. 3b**). With regards to the accurate modeling of the motion of a supercavitating NP, due to the low viscosity, if using a Langevin approach to the driven Brownian motion, it would be essential to retain the inertial terms in the Langevin equations to capture particle dynamics.[43] The experimental observation of such ballistic Brownian motion of supercavitating NPs has not been made due to the difficulty of keeping the NP hot while applying no additional interference to the dynamics due to factors like optical forces.

Plasmonic heating by a pulsed laser can keep an NP hot to maintain the supercavitating state. After a pulse, the supercavitating NP cools down due to the heat diffusion to the surrounding, and then the nanobubble shrinks and eventually disappears. The nanobubble lifetime is 1-100 ns,[16] but if a subsequent pulse arrives within the lifetime of the nanobubble, the Au NP can be heated again to prevent nanobubble from collapsing. A pulsed laser with a repetition rate > 10 MHz can potentially realize such a situation to keep the supercavitating NP for an extended period of time. In the meantime, the laser also applies optical pressure, which acts as a driving force for the NP to move. As the NP is kept hot due to laser irradiation and can instantaneously evaporate water to extend the nanobubble boundary as it is driven by the laser to move, Lee et al.[10] observed that the supercavitating NP could move a long distance of 0.1 mm with speed (0.1 m/s) that is almost 100 times faster than a bare NP in water if driven by the same optical force (**Fig. 3c**). Interestingly, Lee et al.[44] also found that at certain NP-nanobubble geometrical configurations, the photon stream reflected by the water/bubble interface can result in a net optical force on NP opposite to the light propagation direction. They also observed the super-fast backward motion of NPs in the experiment.[10] The supercavitating NP is a thermally non-equilibrium state where continuous photo-thermal energy conversion, heat transfer, water evaporation, and condensation coexist in nanoscale. The above picture is based on plausible physical reasoning, and direct experimental observation can be challenging. Fu et al.[29] used a 4-D TEM to directly visualize NP motion with plasmonic laser excitation. They found that plasmonic nanobubbles can attach to the surface of NPs and propel them to move. Due to the random orientation of the nanobubble and NP, they observed forced Brownian motion (*i.e.*, active Brownian motion) with instantaneous speed up to ~6 m/s. It is noted that this active Brownian NP movement at the nanoscale is fundamentally different from the light-driven ballistic supercavitating NP movement, which is driven by an optical force and always align with the light axis during the ~ 100 micrometer travel distance.

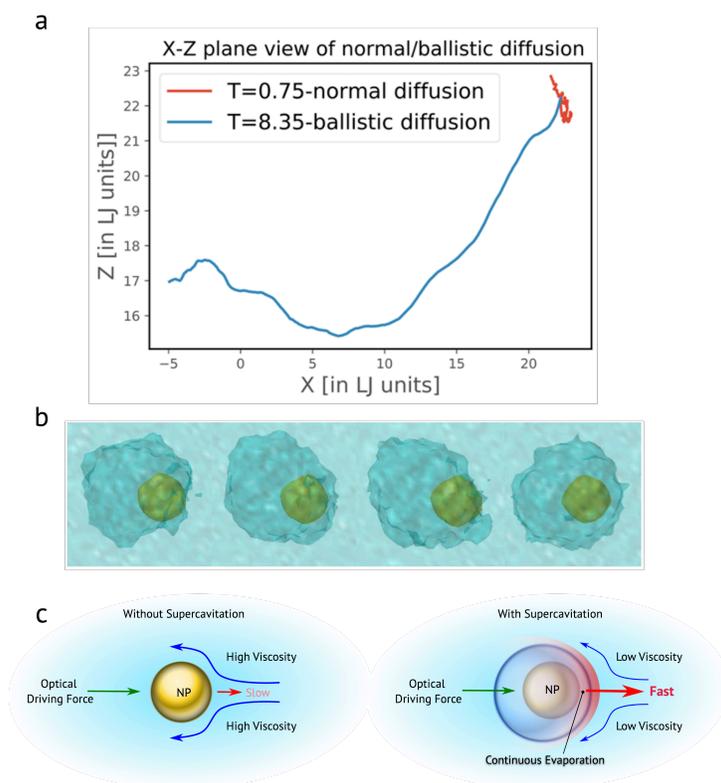

**Figure 3.** (a) Representative trajectory of Brownian motion of NP (*T* = 0.75 in LJ unit) and supercavitating NP (*T* = 8.35 in LJ unit) showing drastically different displacement in the same period of time. (b) MD simulation snapshots of the ballistic Brownian motion of an intensely heated supercavitating NP where the nanobubble boundary is extended as the NP moves due to Leidenfrost effect. (c) Schematic of light-driven NP with and without supercavitation.

**Theoretical modeling:** A wide array of theoretical and simulation approaches have been applied to the study of bubble nucleation and dynamics, ranging from continuum theories of varying complexity[18, 19, 24, 38, 40, 45] to MD simulations.[10, 22, 37, 38, 46, 47] The classical Rayleigh-Plesset equation has been used to model bubble dynamics and collapse.[18, 40, 45] This approach is very successful in modeling bubble dynamics following nucleation of the vapor phase. For example, Wang et al.[45] and Zaytsev et al.[40] used complementary experiments and calculations based on the Rayleigh-Plesset equation to find that dissolved gas in the fluid has a profound effect on bubble collapse dynamics at long times, and that the history of bubble dynamics can influence gas adsorption at the liquid-vapor interface. Such approaches are very effective in understanding bubble dynamics, but do not account for phase change or the role of NP-fluid thermodynamics. On the other hand, theoretical studies using constitutive models within a phase-field like (so-called dynamic Van der Waals) approach[19, 24] can successfully capture phase change and subsequent bubble dynamics in good agreement with the Rayleigh-Plesset description at intermediate-to-long times. Moreover, these models can readily include multi-physics phenomena including NP thermodynamics[19] and solid-fluid interactions.[24] The main advantage to both constitutive hydrodynamic models and the Rayleigh-Plesset equation is that the continuum description can be applied to NPs of arbitrary size for long times at essentially no increase in computational cost for larger NPs. Additionally, the continuum description offers a relatively intuitive understanding of the underlying physics. However, these phase-field/dynamic van der Waals hydrodynamic approaches have relied on simplified fluid models for a single species of molecule and largely ignored the role of dissolved gas and other solutes, which would necessitate introducing multiple gradient-density coupling constants into the theory.

Besides constitutive models, MD simulations have been a useful tool for exploring the fundamental physics related to plasmonic nanobubbles as it can faithfully capture all relevant physics associated with nanobubbles within the simulated model. Many of the above-discussed physics were studied using MD simulations. For example, Sasikumar et al.[22, 38] conducted MD simulations to study the cavitation dynamics around intensely heated solid NPs immersed in a model Lennard-Jones fluid and observed four stages of bubble asymmetric temporal evolution, providing a detailed understanding of the thermal characteristics during the formation and collapse. MD simulations have also shed light on nanobubble dynamics in multi-component fluid. Maheshwari et al.[46] studied the formation of a nanobubble around a heated NP in a model liquid with different concentrations of dissolved gas using MD simulations. They found that beyond a certain threshold concentration, the dissolved gas dramatically facilitated vapor bubble nucleation due to the formation of gaseous weak spots in the liquid surrounding the NP. MD simulations also provide convenience for parametric studies. For example, Pu et al.[47] found that nanobubbles around heated NPs are generated faster if the NP surface is superhydrophobic than hydrophilic. MD has also helped understand the ballistic movements of supercavitating NPs and revealed the nanoscale Leidenfrost effect.[10, 37]

Ideally, MD can include all the complex factors in realistic nanobubbles dynamics around heated NPs, if the force field used in such simulations are accurate and simulation sizes are computationally affordable. However, this is a big "if" since simulating real NPs of tens of nm in size in addition to a sizeable solvent box can be extremely computationally expensive, not to mention the uncertainty related to force fields for the solvent and between NP and solvent. For example, a simulation of a gold NP immersed in water found no bubble formation even if the NP was heated to ~900 K,[48] but simulations of model systems of heated solid NPs in argon showed robust nanobubble formation.[22, 37] MD simulations may also be an integral part of multiscale models by providing important input parameters (e.g., interfacial energy, thermal boundary conductance) or thermodynamics equation of state to mesoscale models, but such a promise is still to be filled. It is also desirable that large-scale molecular simulations of realistic water nanobubble around NP are more commonly performed instead of just toy models. One promising route to addressing all these issues is the incorporation of machine learning methods, which has already been successfully applied to the study of multi-scale cavitation in bulk fluid.[49]

**Applications of nanobubbles:** In last few decades, the plasmonic nanobubble has been investigated mainly for biomedical applications, such as cell-level therapy[3, 50-52] and imaging,[3, 50, 53-56] controlled drug release and delivery [a,b,i-k],[1, 3, 50, 57] and microtissue surgery.[3, 50, 58] The plasmonic photothermal imaging is based on the local variation of refractive index induced by vapor nanobubble. For instance, Lukianova-Helb et al.[54] could selectively insert core-shell Au NPs with a SPR wavelength of 760 nm in target cells (leukemia cells, lung and squamous carcinoma cancer cells), and have confirmed that the NPs form small clusters in the cells. When a pulsed laser tuned at the SPR peak wavelength illuminates these NP clusters, plasmonic nanobubbles were formed. They used these plasmonic nanobubbles as optical amplifiers that increases the scattered intensity of probe light (532 nm) by up to 1800-times compared to that with bare NPs without bubbles. The plasmonic nanobubbles were believed to be an effective way for cell-level imaging without detectable damage to host cells if the laser excitation energy is properly controlled. Upon stronger excitation, the expansion of plasmonic nanobubbles can also induce a mechanical shock that can open up the cellular membrane or open an injected liposome to release drug. These processes can kill target cancer cells [2, 59] or achieve intracellular drug delivery and release.[1]

In biomedical applications, the excitation laser should have a frequency in the biological transparent window (650 – 1350 nm).[60] While solid Au NPs have been a good model system for understanding fundamental physics, their excitation peak would be in the wavelength range where biological tissues are not transparent. Such practical constraints have promoted the design of NPs with suitable SPR frequencies while minimizing fluence threshold for nanobubble generation. At the near-infrared wavelength of ~800 nm, which is transparent to most biological tissues, Lachaine et al.[30] have proposed a design rule of silica-core-Au-shell (core-shell, CS) NP to consider plasmonic nanobubble threshold fluence and irreversible cell damage. They have found that an optimal CS NP (42 nm silica core and 29 nm Au shell) can have a threshold fluence of 350 $Jm^{-2}$ with off-SPR excitation (pink upward triangle in **Fig. 1b**), where 80 % of the CS NPs will not be damaged by overheating after forming nanobubbles. They have also found that the sub-optimal CS NP (cyan pentagram in **Fig. 1b**), which is easy to be damaged due to heating but can have a much lower threshold fluence of 70 $Jm^{-2}$. It is noted that this threshold fluence is much lower than the predicted value of 200 $Jm^{-2}$, but such a difference was not explained. Ogunyankin et al.[28] have proposed hollow Au shell NPs, achieving a threshold fluence of 20 Jm$^{-2}$ with a diameter of 30 nm (gold pentagram in **Fig. 1b**, $\tau$ = 28 ps, $\lambda$ = 800 nm: SPR). Notably, even they used a picosecond pulsed layer, the recorded threshold fluence is very close to Metwally's prediction with femtosecond pulsed laser. The nanobubble formation, however, can severely damage the hollow shell Au NP, melting it into a solid sphere.

Plasmonic nanobubbles have also been explored for vapor generation[6, 7, 61, 62] and plasmon-assisted photocatalytic reactions.[8] Neumann et al.[6] dispersed core-shell NPs in water, and when irradiated by sun light, it was believed that the temperature around NPs was sufficiently high to enable local vapor formation (*i.e.*, nanobubble). They believe these vapor nanobubbles will coalesce with each other under continued illumination leading to larger vapor bubble, which can float towards the water-air interface and then release the vapor (**Fig. 4a**). However, it was later concluded that nanobubble formation was impossible given the low solar intensity,[61] and the observed water evaporation would be evaporation at the water-air interface. A similar strategy was employed for plasmon-assisted photocatalytic reactions. Adleman et al.[8] used plasmonic NPs as heterogeneous catalysis to provide both heat and vaporized reactants to the system without excessive heating of the immediate surroundings or the preheating needed to vaporize the reactants.

As previously mentioned, the supercavitating NPs can be driven by light. Using this feature, Zhang et al.[63] used light to deposit Au NPs onto the transparent surface by optical forces. The deposited Au NPs become the surface heater triggering the nucleation of micro-sized plasmonic surface bubble (**Fig. 4b**).[63, 64] The deposited NPs, when heated by a laser, can also help de-pin the front contact line of the plasmonic surface bubble, enabling light to guide the surface bubble to move (**Fig. 4c** left panel).[11] As the surface bubble moves with the light, NPs can be deposited along the moving path. This contact-line deposition can be used as a controllable technique for pattern writing (**Fig. 4c** right panel).[12, 65] The formed plasmonic surface bubble has also been leveraged for bio-sensing. The Marangoni flow around the surface bubble helps bring analytes in the solution to the bubble surface,[66, 67] which then deposits the analytes to the contact line of the surface bubble. Leveraging the shrinking phase of the bubble, Moon et al.[5] deposited and concentrated hairpin DNA-functionalized Au NPs on a substrate which significantly reduced the detection limit using fluorescence signals (**Fig. 4d**).

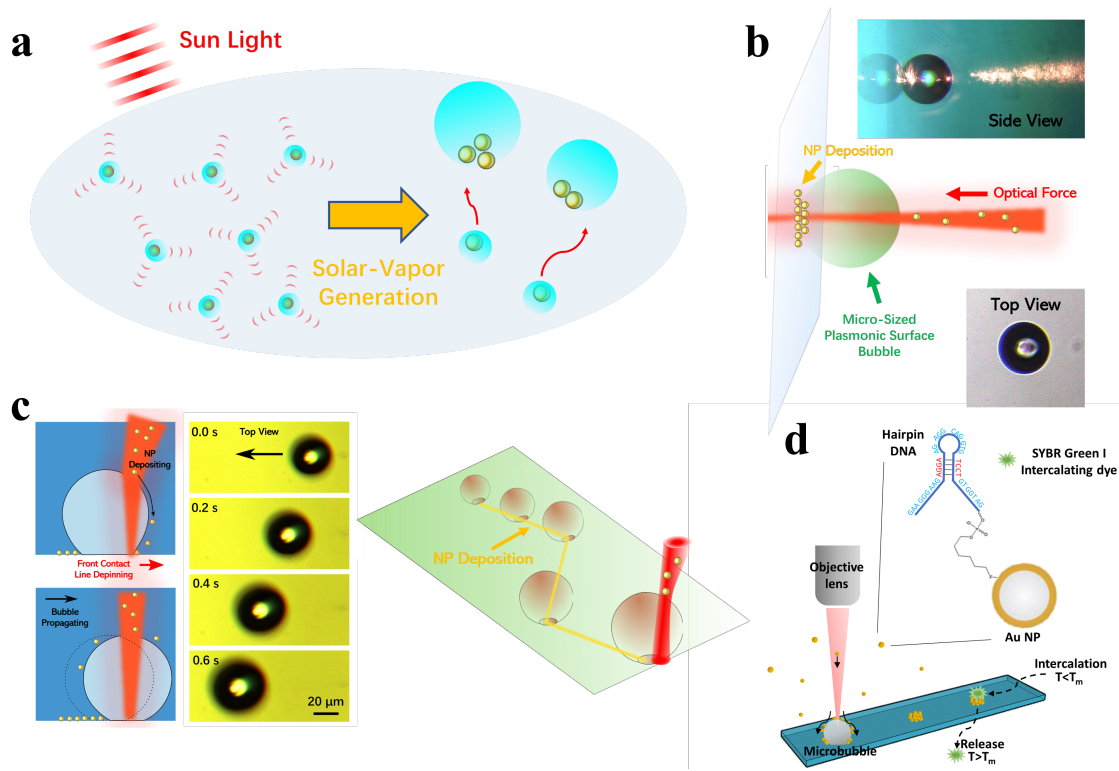

**Figure 4**. (a) Schematic of the originally believed NP-enabled solar steam generation.[6] (b) Schematic and optical images of the plasmonic surface bubble generated by the optically deposited Au NPs. (c) The schematic and optical images showing the optically directed surface bubble movement and NP deposition process on transparent surfaces. (d) Schematics of the hairpin DNA-functionalized Au NPs deposition and concentration by leveraging the shrinking phase of surface plasmonic bubble.

Bio-sensing applications using plasmonic bubbles are still developing, but most of such applications leverage the flow surrounding the plasmonic surface bubble. This flow has been called "bubble tweezer" and can be mainly induced by the difference of bubble/liquid interfacial tension (*i.e.*, Marangoni flow), which depends on temperature gradients. These different bio-sensing applications leverage similar strategies in generating the surface bubble, where they use the photo-thermal conversion process occurring in 3D patterns,[68] optically resistive thin-films[69] or metallic NPs.[5, 70] The bubble tweezer can actively captures analytes in the solution and bring them to the three-phase contact lines of the surface bubble. The deposited analytes are later detected via techniques like surface enhanced Raman spectroscopy (SERS) or fluorescence. On the other hand, using the bubble tweezer can cause the deposited analytes to have a higher temperature, which may deteriorate the function of the analytes. Moon et al.[5] have shown that such heating problems can be avoided when leveraging the shrinkage of the surface bubble after the laser excitation is turned off. In the bubble-shrinking process, the analytes trapped at the surface bubble are eventually deposited at the contact lines without heating. We have summarized the studies in bio-sensing applications in **Table 1**, which shows that the limit of detection (LOD) can be as low as 10 fM. The LOD is found inversely correlated with the bubble size.[5] It suggests that larger bubbles would be preferred to lower the LOD. However, it may worsen the damage of analytes by the photo-thermal heating effect or pose a longer shrinkage time (~hours) when we use the shrinkage bubble.[5] Thus, further research is needed to continue to decrease LOD so that the bubble-assisted active sensing technology can be used for biomarker detections such as influenza and cancer, which require LOD in the femto- or atto-molar level.[71-73]

Table 1. Active bio-sensing applications using plasmonic surface bubbles

| Plasmonic bubble generation | Laser wavelength (power density) | Bubble size (Shrinking / concentrating time) | Strategy for concentrating analytes | Target | Detection method | LOD |
|---|---|---|---|---|---|---|
| Functionalized Au nanoparticle suspension[5] | 800 nm (0.88 mW/$\mu m^2$) | ~ 40 μm (~7 min) | Shrinking bubble deposition | Hairpin DNA-functionalized Au NPs | Fluorescence | - |

| 3D nanoantennas[68] | 850 nm (390 mW/μm$^2$) | ~ 100 μm (~ 2 min) | Shrinking bubble deposition | Extracellular membrane vesicles (Evs) | Surface enhanced Raman spectroscopy (SERS) | - |
|---|---|---|---|---|---|---|
| Array of Au nanoislands / Perfluoropentane (PFP)[74] | 532 nm (0.26 mW/μm$^2$) | ~ 20 μm (1 min) | Contact line deposition | FITC-Protein A/G | Fluorescence | 10 nM |
| Moiré Chiral Metamaterials [75] | 532 nm (-) | ~ 5 μm (20 min) | Successive microbubble shrinking deposition | Glucose | Circular dichroism | 100 pM |
| Au nanoisland film[69] | 785 nm (0.2 mW/μm$^2$) | ~ 100 μm (~ tens of min) | Shrinking bubble deposition | 4-MBA / R6G | SERS | 1 pM / 100 nM |
| Accumulation of Au nanoparticles[70] (Nanosphere / Nanoshell) | 1064 nm (10 mW/μm$^2$) | ~ 175 μm (2 min) | Contact line deposition | R6G / Malachite Green fungicide | SERS (785 nm) | 10 fM |

## 3. Summary

In this perspective, we have reviewed the current state of the understanding of the nanobubble formation and dynamics physics, the theoretical modeling effort to describe the multi-scale, multi-physics phenomena, and the applications. The minimal threshold fluence for nanobubble formation has been qualitatively described by heat transfer models, but factors like the nature of laser excitation (inter-band vs. SPR) and temperature-dependent optical absorption efficiency need to be considered to improve the model so that their prediction can be closer to experimental data. For more detailed theoretical models, while they have been able to explain important physics underlying the fluence threshold such as NP thermodynamics, interfacial thermal conductance, dissipation vs. phase change, more accurate thermodynamics equation of the fluid surrounding the intensely heated NP and the treatment of multi-component fluid (e.g., dissolved air in water) are the next steps to enhance the capability of these models. MD simulations have been an important tool to provide molecular level understanding of nanobubbles, but most of them are on toy models as they are limited by the computational length and time scales needed to simulate more realistic systems. It is also noted that so far, no MD simulation has been able to simulate nanobubble formation of NP immersed in water, even if in the system size current computation can handle. It is desired that MD simulations can provide key information (e.g., interfacial thermal conductance, equation of state) for mesoscale constitutive models to achieve more accurate description of nanobubble formation and dynamics. Plasmonic nanobubbles have also seen a number of potential applications, most of them are bio-medical related like cell-level therapy and imaging, controlled drug release and delivery, and microtissue surgery. For these applications, it is important that the energy of the excitation laser can be minimized so as to reduce side effects from the photothermal effect, and thus we have seen effort to design NPs that can lower the fluence threshold for plasmonic nanobubble formation. NP deposited on a surface can lead to plasmonic surface bubbles, which also have wide active bio-sensing applications, but further research is needed to lower the sensing LOD so that they can be practiced to for more challenging detections like influenza and cancer.

Overall, plasmonic nanobubble is a phenomenon that involve complicated and intertwined optical, thermal and fluidic physics, which needs multi-disciplinary effort to fully understand. New multiscale simulation models need to be developed to accurately describe this phenomenon, but there are challenges to include all the complicated physics into one model. Most of the experimental understandings are based on inference from microscopic observations. Experimental tools that resolve the time and spatial scales of nanobubbles are needed to better validate modeling results, and high-speed TEM has emerged as a promising tool. Applications may benefit from such fundamental understandings in aspect like optimizing the laser excitation conditions and NP designs.

**Acknowledgement**

The authors thank support from multiple National Science Foundation grants (1706039, 1931850, 2040565 and 2001079) and the Center for the Advancement of Science in Space (GA-2018-268). E.L. would like to also thank the supported from the National Research Foundation of Korea (NRF) grant funded by the Korea government (MSIT) (No. NRF-2021R1C1C1006251).